**Utility of ocean wave parameters in ambient noise prediction**

W. Erick Rogers,[1a] Laurie T. Fialkowski[2], Daniel J. Brooker[2], Gleb Panteleev,[1] and Joseph M. Fialkowski[2,3]

[1] *Ocean Sciences Division, Naval Research Laboratory, Stennis Space Center, MS, 39529, USA*

[2] *Acoustics Division, Naval Research Laboratory, Washington, DC, 20375, USA*

[3] *Retired*

Abstract. This study is concerned with prediction of the "wind noise" component of ambient noise (AN) in the ocean. It builds on the seminal paper by Felizardo and Melville (1995), in which the authors quantified the correlation between AN and individual wind/wave parameters. Acoustic data are obtained from hydrophones at six diverse locations, and wind/wave parameters are obtained from moored buoys and numerical models. We describe a procedure developed for this study which identifies correlation of AN with wave parameters, independent of their mutual correlation with wind speed. We then describe paired calibration/prediction experiments, whereby multiple wind/wave parameters are used simultaneously to estimate AN. We find that the improvement from inclusion of wave parameters is robust but marginal: typically RMSE is reduced by less than 0.3 dB and/or less than 12% of the original RMSE. We interpret the latter outcome as suggesting that wave breaking responds to changes in local winds quickly, relative to, for example, total wave energy, which develops more slowly. This outcome is consistent with prior knowledge of the physics of wave breaking, e.g. Babanin (2011). We discuss this in context of the time/space response of various wave parameters to wind forcing.

---

[a] Email: w.e.rogers.civ@us.navy.mil

# I.    INTRODUCTION

Ambient noise in the ocean is produced by shipping, biologics, wind, precipitation, sea ice, and other sources. The goal of this study is to improve prediction of the component of ocean ambient noise (AN) commonly referred to as "wind noise". Important early work on this topic was performed by Knudsen et al. (1948) and Wenz (1962). In these works, it was recognized that the sea state ("surface agitation") has a central role in the creation of this noise, while the underlying mechanisms were imprecisely ascribed to some combination of the action of bubbles in the water and/or the impact of sea spray on the surface. Wilson (1980) argued that the latter was a dominant mechanism, but by the end of the decade, there was general recognition that bubbles play the more important role (e.g., Prosperetti (1988), Medwin and Daniel (1990)). Prosperetti (1988) outlined three frequency regimes: 1) low frequencies, 100 to 200 Hz, where bubbles in the water act as amplifiers of "water turbulence noise", 2) intermediate frequencies, 200 Hz to 1 or 2 kHz, where there is typically a broad maximum in the wind/wave noise, caused by "collective oscillation of bubble clouds", and 3) high frequencies, 1 or 2 kHz to around 10 kHz, where individual bubble oscillation dominates. The last regime is the most well understood: this noise is emitted immediately (e.g. 10s of milliseconds, Manasseh et al. 2006) after the bubbles are formed, and the frequency of the emission is directly related to the bubble radius (Minnaert 1933; Manasseh et al. 2006).

It is generally understood that wave breaking is the link between wind speed and bubble generation. Further, Farmer and Vagle (1988) established that the third noise mechanism above is a direct result of wave breaking. However, in the real ocean, wave breaking is not a simple function of wind speed. While wind speed is of primary importance, the frequency and severity of wave breaking is affected by other environmental factors, e.g., see review by Babanin (2011). Consider, as illustration, an idealized case of 15 m/s wind speed over 100 m basin vs. the same wind over 100 km basin. The oceanographic difference (specifically, fetch) will result in waves of different height and length, different strength of breaking, different bubble generation, and therefore different acoustic emissions for these two cases. Current AN models use wind speed, plus water depth (usually binary, i.e. deep vs. shallow, Wenz (1962)) and/or receiver depth Hildebrand et al. (2021) and predict AN as a function of frequency.  Our hypothesis is that since the AN is more directly connected to wave breaking than to wind speed, then it should be possible to improve a prediction of wind noise by incorporating oceanographic information, particularly that associated with ocean waves.

The above hypothesis is not new. It was central to the landmark study of Felizardo and Melville (1995) (henceforth denoted "FM95"), who quantified the correlation between AN and several wind/wave parameters for a relatively short (10-day) field experiment, looking at three frequencies, from 4.3 to 14 kHz. They found that the AN correlation with predicted wave dissipation rate is equal to, or slightly higher, than the correlation with wind speed. In the present study, we perform a similar analysis, though here we apply multiple input parameters simultaneously to predict AN, rather than one parameter at a time as in FM95. Further, we use a larger dataset: the cumulative duration of our dataset is 1291 days, making it more than 100 times larger than that of FM95, and a much broader range of frequencies are studied (127 Hz - 18 kHz, vs. 4.3 – 14 kHz). We employ wind and wave information at seven different hydrophone locations. At one location, wave characteristics are taken from observations, while at the other six locations, wave parameters are derived from numerical ocean wave models.

In Section II, we develop further our thought experiment of idealized wave growth, to illustrate the uncertainty around the behavior of emitted AN as a type of "wave parameter". In Section III, we introduce the datasets used in the study. In Section IV, we present an analysis in which we study the correlation of AN with wave parameters, after the correlation with wind speed is constrained. In Section V we perform multi-linear regression to predict AN using combined wind and wave parameters and quantify the benefit of including wave parameters in terms of RMS error in dB, relative to the baseline performance, using wind information alone. In Section VI, we discuss limitations and future work, and in Section VII, we summarize our conclusions from this study.

Since Wenz (1962), empirical predictions of AN using wind speed have demonstrated good skill. For example, analysis by Yang et al. (2023, their figure 2(c)), finds that 80% of the data are within 3 dB of the mean for 6-8 m/s wind speed at 5 kHz. The present study seeks to provide quantitative information on the potential for inclusion of wave information to improve this skill. We do not present a new predictive formula as was done, for example, by Wenz (1962), Hildebrand et al. (2021), and Pflug (2021). Also, bubbles are not explicitly part of the predictive models developed herein. The problem of noise from breaking waves in the surf zone, e.g. see Deane (1997) and Fabre and Wilson (1997), is not addressed here.

## II. IDEALIZED CASE: FETCH-DEPENDENCE

In Section 1, we introduced a thought experiment with 15 m/s wind speed blowing over two basins, one with 100 m fetch, and the other 100 km, and stated that the latter case will logically result in more bubbles and ambient noise. We elaborate further here. According to prevailing theory, the wave spectrum is controlled by the net effect of three "source terms" which dominate in deep, open water: wind input, wave breaking dissipation, and weakly non-linear four-wave interactions (Cavaleri et al. (2007)). The third is a "direct and inverse cascade", meaning a transfer of energy from near the spectral peak to the higher and lower frequencies, respectively (e.g., Young and van Vledder 1993). At higher frequencies, say 1.5 to 3 times the peak frequency, these three source terms roughly balance each other, so that the spectral level is in rough equilibrium with the wind (e.g., Phillips 1985). At frequencies nearer the peak, in growing seas, the local (in spectral space) imbalance of the three terms produces a spectral downshifting with increasing fetch or duration of a wind event: producing larger wave height (thus greater wave energy) and longer dominant wave lengths, e.g., Kahma (1981). It might be understood intuitively that breaking in more mature or 'fully arisen' seas produces more turbulence, bubbles, and AN than breaking in less energetic windsea, but it is also supported in the literature. Using observational data, Zhao and Toba (2001), in their Fig. 1 show whitecap coverage increasing with wave age. Using fitted curves based on energy balance equations and observational data, Hwang and Sletten (2008) show in their Figure 3c, parameter $\alpha$ increases with the duration of a wind event, where $\alpha$ is the coefficient of proportionality between total energy dissipation and the wind speed cubed.

The above suggests that, for any particular wind speed, AN should grow with fetch. However, this is merely a qualitative relation, and so it does not assist us in predicting AN. A numerical ocean wave model such as SWAN (Simulating WAves Nearshore, Booij et al. (1999)) is based on the energy balance described above and can be used to predict many different wave parameters, each of which is associated with wave breaking to a greater or lesser degree. Table 1 shows SWAN results for an idealized case with a 10-meter wind speed of 15 m/s.

TABLE I. Output from SWAN model for idealized case of wind speed of 15 m/s. Percentages shown are "percent relative to fetch of 300 km".

| fetch length | 300 m | 3 km | 30 km | 150 km |
|---|---|---|---|---|
| fetch (%) | 0.1% | 1.0% | 10% | 50% |
| $H_{sig}^{0.25}$ | 46% | 61% | 83% | 95% |
| $H_{sig}^{0.5}$ | 21% | 37% | 69% | 91% |

| $H_{sig}$ | 4.6% | 14% | 48% | 83% |
|---|---|---|---|---|
| $T_p$ | 18% | 35% | 62% | 91% |
| $D_{tot}$ | 15% | 40% | 96% | 100% |
| $m_{-1}$ | 0.0% | 0.6% | 14% | 60% |
| $m_0$ | 0.2% | 1.9% | 23% | 69% |
| $m_1$ | 0.9% | 5.7% | 35% | 78% |
| $m_2$ | 3.9% | 15% | 53% | 86% |
| $m_3$ | 13% | 35% | 73% | 94% |
| $m_4$ | 32% | 61% | 88% | 98% |
| $m_5$ | 54% | 84% | 96% | 99% |
| AN | unknown | unknown | unknown | unknown |

Parameters given in Table 1 include moments of the wave energy density spectrum $E(f)$. They are computed as $m_n = \int_{f_1}^{f_2} E(f) f^n df$, where $f_1$ and $f_2$ are the upper and lower wave spectra frequency bounds, with 0.038 Hz and 1.0 Hz used here, respectively[1]. Spectral moments typically have a physical relevance. For example, the significant waveheight $H_{sig}$ is defined as $H_{sig} = 4\sqrt{m_0}$; $m_2$ is proportional to surface orbital velocity; $m_3$ is proportional to surface Stokes drift; $m_4$ is proportional to mean square slope of the sea surface. Parameter $D_{tot}$ is the integrated rate of dissipation of wave energy by whitecapping, and $T_p$ is peak period. The calculation and physical relevance of these parameters are described in greater detail in Appendix A.

The wave parameters grow at a markedly different rate. For example, $D_{tot}$, $m_4$ and $m_5$ grow very quickly with fetch, while total energy ($m_0$) grows more slowly. For the "wave parameter" AN generated by breaking waves, the growth rate is unknown, though intuition would suggest a growth rate similar to $D_{tot}$, $m_4$ and $m_5$. And of course, it must be recognized that even this unknown idealized growth rate of AN is likely dependent on the acoustic frequency.

## III.   DATASETS

This study relies on three datasets: 1) hydrophone observations of ambient noise, 2) observed wind/wave parameters, and 3) modeled wind/wave parameters. Each is described below.

### A.  Hydrophones

The hydrophones are described in this section. Below, we refer to critical depth (CD), which is the depth below the sound speed minimum axis where sound speed is equal to the maximum above the axis (see Urick (1983) and Kuperman and Roux (2007)). We briefly review its relevance here. For cases when the ocean depth is below the critical depth, acoustic energy does not interact significantly with the bottom, and long-distance propagation is supported. In cases where the ocean bottom is above the critical depth, sound is more significantly attenuated by the bottom. The amount of ambient acoustic energy reaching a hydrophone is maximized when a receiver is shallower than the critical depth and the ocean bottom is deeper than the critical depth. When both receiver and ocean bottom are shallower than critical depth, then most of the acoustic energy that reaches the receiver has interacted with the ocean bottom and is more attenuated. When the receiver and ocean bottom are both below the critical depth, then the receiver lies in the "shadow zone," where most ambient sound waves are refracted away from the receiver and most received AN is that which is generated locally; an assumption of "direct path only" propagation is appropriate in this case (Wilson 1983).

## 1. NOAA/PMEL

The first hydrophone dataset is from a deployment by the National Oceanic and Atmospheric Administration (NOAA) Pacific Marine Environmental Laboratory (PMEL) at Ocean Station Papa (OSP), south of the Gulf of Alaska. The location is very remote, around 1000 km from the Alaska and Canada mainlands, so fetch here can be limited only by the scale of the meteorological forcing. It is in deep water (4250 m depth), with the hydrophone at 500 m depth, which is above critical depth. Data were processed for October to December 2018. The processing of these data include eight analysis frequency bands, with center frequencies from 127 Hz to 1.7 kHz, with 5 kHz sample rate.

## 2. ONC

The second and third hydrophones are offshore of southwest Canada, deployed and managed by Ocean Networks Canada. Both are near the seafloor, and data were processed for October 2020 to December 2021. One hydrophone is the "Cascadia Basin" deployment, 202 km west of Vancouver Island. The more nearshore hydrophone is the "Clayoquot Slope" deployment, 78 km west of Vancouver Island. The wind climate for this region is primarily alongshore, so fetch-limited conditions are not common. These data include 18 analysis frequency bands, with center frequencies from 127 Hz to 18 kHz, with 64 kHz sample rate. At most frequencies, Cascadia is quieter than Clayoquot, implying that at the latter, the impact of being within the sound channel is relatively more important than any suppression of sound by bottom interaction.

Some of the results herein from the OSP and ONC datasets were previously presented in the grey literature (Rogers et al. 2023).

## 3. HARP

The remaining four datasets are from hydrophone data provided directly by Dr. John Hildebrand (UCSD). These are the High-frequency Acoustic Recording Package (HARP) instruments described in Hildebrand et al. (2021).

## 4. Hydrophone summary

The instrument locations are shown in Fig. 1 and features of the hydrophone dataset are summarized in Table II. This table includes a simple metric for roughly estimating the fraction of each dataset which is fetch- or duration-limited based on the concept of separating between "forced conditions" (i.e., waves actively receiving energy from the wind) and "unforced conditions"[2]. This is denoted as "%$C_r/U < 1$", and here, $U$ is the 10-m wind speed, $C_r$ is an ocean wave phase velocity representative of the wave spectrum, calculated using the deep-water relation, $C_r = gT_{m,-1,0}/(2\pi)$ where $g$ is gravitational acceleration, and $T_{m,-1,0}$ is a definition of mean wave period, $T_{m,-1,0} = m_{-1}/m_0$ (other definitions of mean period exist; see Appendix A). The smaller percentage—implying larger $C_r/U$—at the OSP and ONC sites indicate a more mature sea state, implying that a sea state prediction will be less heavily penalized for using only local wind speed.

---

[2] It must be emphasized that, in reality, a typical wave spectrum is composed of both forced and unforced components, with wind and wave direction playing an important role, implying that the metric $C/U$ is imprecise and is used here only to roughly compare the wave climate at each site. This is discussed further in Section VI.D.

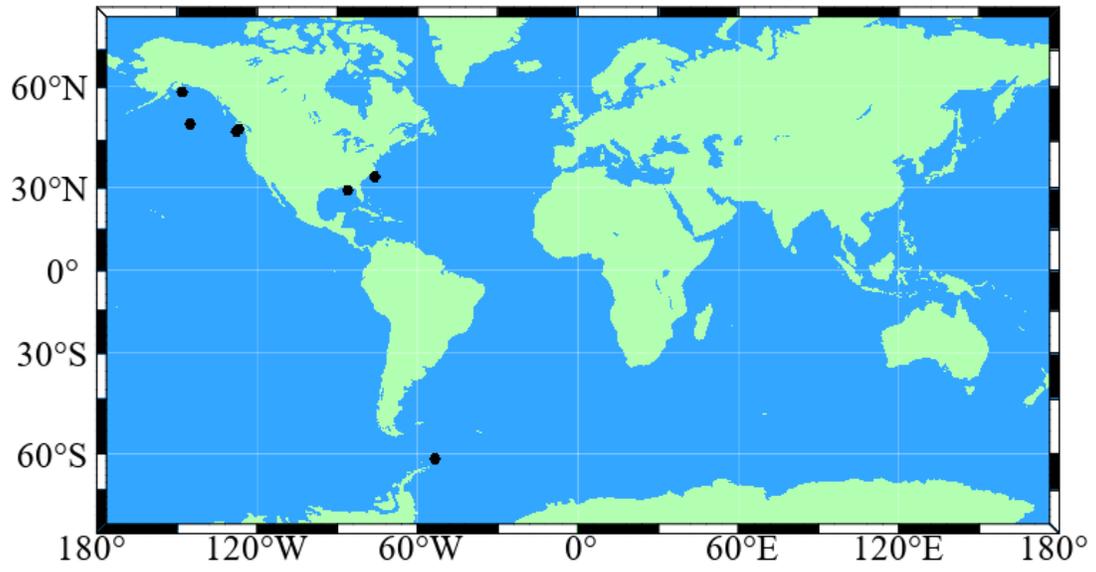

FIG. 1. Positions of hydrophones used in this study.

Table II. Summary of hydrophone data used in this study. "Long description" is, in the case of the HARP buoys, the designation in Hildebrand et al. (2021, their Table V). Except for the OSP location, hydrophones are mounted near the seafloor (water depth is hydrophone depth). The implications of hydrophone placement relative to critical depth (CD) is explained in the text. The last column, "% $C_r/U<1$", is a rough measure for the fraction of records with wave spectra that are actively forced by the local wind (see text).

| Short-hand used herein | Long description | Start and end of data used herein | # of 30-minute co-locations | position | water depth (m) | hydro-phone rel. to critical depth (CD) | % $C_r/U$ <1 |
|---|---|---|---|---|---|---|---|
| OSP | NOAA/PMEL Station NRS02 | 2018/10/01 - 2019/01/01 | 4,403 | 214.87°E, 50.25°N | 4250* | above CD | 20 |
| Casca-dia | ONC-Cascadia Basin | 2020/10/03 - 2021/12/17 | 13,311 | 232.23 °E, 47.77°N | 2660 | below CD | 12 |
| Clayo-quot | ONC-Clayoquot Slope | 2020/10/03 - 2021/12/17 | 10,946 | 233.15°E, 48.67°N | 1260 | no CD | 16 |
| GofAK | HARP GofAK_CB_09 | 2017/09/15 - 2017/12/31 | 5,147 | 211.97°E, 58.67°N | 900 | no CD or at CD** | 20 |
| WAT | HARP WAT_GS_02 | 2018/01/01 - 2018/04/01 | 4,292 | 284.00°E, 33.67°N | 930 | no CD | 46 |
| Antarc | HARP Antarc_EIE_01 | 2016/02/02 - 2016/11/30 | 13,775 | 306.52°E, 61.25°S | 1033 | below CD | 35 |
| GofMX | HARP GofMX_DC_11 | 2017/10/03 - 2018/04/30 | 10,079 | 273.90°E, 29.05°N | 269 | no CD | 56 |

\* OSP hydrophone is 500 m below surface

\*\* This describes May through December only. At GofAK hydrophone location, there is no critical depth (CD) during May through November, and hydrophone is at CD in December.

## 5. Acoustic data processing

The procedure for wind noise estimation is similar for all data sets, with modifications to account for factors such as data bandwidth, hydrophone depth, and acoustic propagation characteristics. Raw hydrophone time series data were first processed using Welch's method (Welch 1967) to produce spectral estimates over stationary time intervals. A Hann window was used to minimize off-frequency-bin energy leakage via spectral sidelobes (Harris 1978). Received sound levels were then estimated over analysis frequency bands spanning 100 Hz to 20 kHz with center frequencies and bin widths chosen to avoid redundancy. Analysis bin center frequencies for the 18 (eight in the case of OSP, due to lower sampling rate) ambient noise analysis bands are shown in Table III. For all data sets, lower frequencies (< 100 Hz) were not processed due to elevated ambient sound levels from shipping activity. High frequency processing cutoffs were determined based on identified low pass filter roll-off and data acquisition system noise floor. Additional description of the quality control process used on the hydrophone data is given in Appendix B.

Table III. Acoustic center frequencies and bandwidths for analysis. Bins 1 − 8 were used for the PMEL data while bins 1 − 18 were used for all other hydrophone data used herein.

| Bin # | 1 | 2 | 3 | 4 | 5 | 6 | 7 | 8 | 9 |
|-------|------|------|------|------|------|------|------|------|------|
| Fc (Hz) | 127 | 191 | 281 | 408 | 587 | 841 | 1199 | 1704 | 2260 |
| ΔF (Hz) | 53 | 75 | 105 | 149 | 210 | 297 | 419 | 592 | 520 |
| Bin # | 10 | 11 | 12 | 13 | 14 | 15 | 16 | 17 | 18 |
| Fc (Hz) | 2848 | 3588 | 4700 | 5875 | 7175 | 9040 | 11390 | 14350 | 18080 |
| ΔF (Hz) | 655 | 855 | 1400 | 950 | 1650 | 2080 | 2620 | 3300 | 4160 |

### B.  Observed wind/wave parameters

The OSP hydrophone is co-located with a PMEL buoy equipped with an anemometer at 4 m elevation (Ocean Climate Station (OCS) Project Office), and an Applied Physics Laboratory / University of Washington wave buoy (Thomson et al., 2015), which is Coastal Data Information Program (CDIP) Station 166[3].

The ONC hydrophones are not co-located with wind/wave observations, but a number of U.S. and Canadian buoys are in the region, eight of which are used to validate input/output to/from the 'SW (southwest) Canada' wave model described in the next section. The HARP locations similarly have two to five buoys in each area which were used for wave model validation, with the exception of the 'Antarc' hydrophone, for which no contemporaneous/nearby in situ wave observations exist.

### C.  Modeled wind/wave parameters

Wind and wave information from numerical models are generally less accurate than information taken from direct observations. However, in an operational environment, observations are not available at all locations, and one must rely on output from numerical models. This is especially true when a forecast is needed.

For all locations other than the 'GofMX', we use a regional grid of the SWAN wave model ("Simulating WAves Nearshore", Booij et al. (1999)) nested within a global 1/4° hindcast with the wave model WAVEWATCH III® ("WW3", Tolman (1991), WW3DG (2019)).  The grid design of Rogers and Linzell (2018) is used for the global grid and resolutions from 1.8 to 2.8 km are used in the SWAN grids. As an example, the SW Canada (ONC) case, the resolution of the regional model is between 2.5 and 2.8 km and is illustrated in Fig. 2. For the 'GofMX' case, a global model was not needed; instead, a 2.1 km × 2.2 km coastal SWAN grid was nested within a 11.9 km × 13.9 km regional SWAN grid. The model settings and SWAN grid information are described in the Supplementary Material.

In the case of the two ONC locations, 10-m wind vector forcing fields for the models comes from the Navy's operational global atmospheric model, NAVGEM (Hogan et al. 2014), given at 0.18° resolution. The global wave model used to provide boundary conditions also includes ice fraction information taken from NAVGEM.  In the case of the HARP locations, wind and ice forcing is taken from the European Centre for Medium-Range Weather Forecast (ECMWF) Reanalysis v5 ("ERA5", Hersbach et al., 2020), which are given hourly at 1/4° resolution, and the

---

10-m neutral winds are used[4]. Model results for the OSP location are not included in this manuscript[5]: only the observed wave parameters are used for OSP herein.

As mentioned in Section B above, with the exception of the 'Antarc' case, the SWAN implementations are validated using wave observations from moored buoys. These buoys are listed in the Supplemental Material.

WW3 and SWAN are both phase-averaged wave models which take inputs such as 10-m wind vectors and use them to predict directional wave spectral density, as a function of space, time, frequency, and direction. From the spectra, wave parameters such as the moments defined in Section II can be derived. They account for the effects of spatially and temporally varying winds (i.e. increasing, falling, turning) on wave growth, dissipation effects such as that by breaking, bottom friction, and interaction with sea ice, and advection in all dimensions. This makes them much more sophisticated (and complex) than a parametric model based on time/space averaged wind speed and approximated fetch. The model's source terms can also be integrated to compute quantities such as energy fluxes and momentum fluxes. In this paper, we utilize $D_{tot}$, which is the frequency- and direction- integrated value of the spectral dissipation by whitecapping, $S_{ds}(f,\theta)$, which is the primary component of energy flux from the waves to the ocean:

$$D_{tot} = \int_{f_1}^{f_2} \int_{0}^{2\pi} S_{ds}(f,\theta)df\,d\theta \qquad (1)$$

For further information on the wave model governing equations, source terms, and methods, the reader is referred to Tolman (1991), Tolman and Booij (1998), Booij et al. (1999), WW3DG (2019), and SWAN (2019).

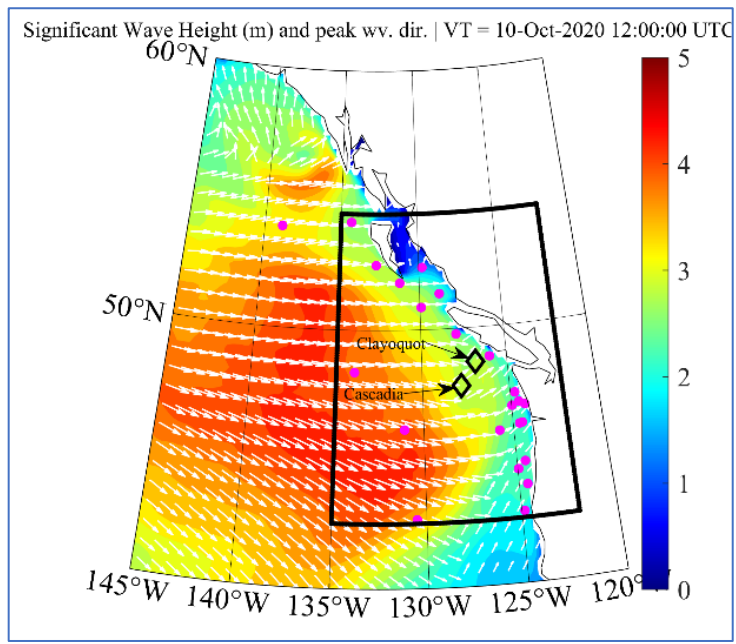

FIG. 2. Significant wave height (colors, in meters) and peak wave direction (white arrows), from the SW Canada SWAN model. Valid time is 2020 Oct. 10 12:00 UTC. The ONC hydrophones are

---

[4] 10-m neutral winds are generally preferred over standard 10-m winds for forcing wave models, but are not available from some atmospheric models, such as NAVGEM.

[5] Both wave models were, in fact, run for the OSP location, using both the NAVGEM and ERA5 forcing fields. The results are not presented here, for reasons explained in Section V. However, the uniquely co-located wave observations make the OSP case too valuable to exclude altogether.

indicated with black diamonds. Magenta dots indicate metocean buoys deployed by National Data Buoy Center (NDBC) of the National Oceanic and Atmospheric Agency (NOAA), and by 'Environment and Climate Change Canada' (ECCC). The open edges of the grid receive boundary forcing from WW3.

### D. Co-locations

In the case of OSP observed wind/wave parameters, the hydrophone data are already geographically co-located. For other cases, i.e., those that employ *model* wind/wave parameters, bilinear interpolation is used for geographic co-location.

Temporal co-locations are made by taking the wind/wave parameters at 30-minute intervals, and for each temporal data point, averaging all valid AN data within ±15-minutes of the wind/wave time. From this, there are between 4000 and 14000 30-minute co-locations for each hydrophone, as listed in Table II. Across all locations, this results in a total of 1291 days of co-locations. Of the co-locations in this total, some include frequencies for which the AN data are marked as invalid by the quality control. The average fraction of valid AN data was 68% for the OSP case, 86-93% for the two ONC cases, and 95-98% for the four HARP cases.

### E. Wind climatology at hydrophone locations

Waves that are not fully developed are either fetch- or duration-limited, which may be caused by non-uniformity and non-stationarity (respectively) of the winds. In the classic scenario, fetch is determined by the distance from a nearby coastline. Wind climatology can help us to understand how likely this scenario is, for each of our hydrophone locations, so here we briefly review this, using information taken from the "GLOSWAC" (GLObal Spectral WAve Climate) database of Portilla-Yandún (2018). All else being equal, fetch- or duration-constrained (i.e. "young") wave conditions are more likely to occur when wind speed is high; the "wave age" is typically defined as $C/(U cos(\theta))$, where $C$ is wave phase velocity (proportional to wave period if in deep water), $U$ is the wind speed, and $\theta$ is the angle between wind and wave directions. Below, for brevity, we use shorthand notation for cardinal directions, e.g. "WSW" for winds from the west-southwest.

Ocean Station Papa (OSP, NOAA/PMEL): Westerly (especially WSW) winds are most common. Winds are often strong ($U_{10}$>12 m/s), and occasionally reach 20 m/s.

Cascadia and Clayoquot (ONC): Winds are primarily along the coast, with winds from NW being most common but with the highest winds occurring more often from the SE.

Gulf of Alaska ('GofAK'): Winds from ESE and W are most common, and winds from north are least common. In the case of westerly winds, there is potential for limiting of wave conditions by fetch from the Alaska coastline, but since the distance is quite large (225 to 300 km), waves are expected to be fetch-constrained only for the strongest winds (e.g. $U_{10}$>15 m/s).

Offshore of North Carolina ('WAT'): Similar to the ONC location, winds are often along the coast, with SW winds being most common.

Drake Passage ('Antarc'): Westerly winds are by far the most common. Winds are strong, occasionally reaching 20 m/s. Clarence Island is only 30 km west of the hydrophone, which strongly implies fetch-limitations, but the island is small, so it would have reduced effect on winds from NW and SW.

Gulf of Mexico ('GofMX'): Winds from E and SE are most common, though northerly winds are also common. Winds typically do not exceed 16 m/s. The coast is 120 to 150 km from this hydrophone. Waves are expected to be fetch-constrained only weakly, and only for the highest winds speeds.

Based on above, only the 'Antarc' location is likely to be fetch-constrained by topography for a significant fraction of the time series used. At other locations, the wave conditions may still be constrained (i.e., not fully-developed), but this would occur primarily as a result of non-uniformity and/or non-stationarity of the wind speed and direction.

## IV. DE-CORRELATION ANALYSIS

Ambient noise is created by bubbles, which are created by breaking waves, and the wave energy is generated by wind action. Thus, the AN is strongly correlated with wind speed, and so models such as Wenz (1962) and Hildebrand et al. (2021) have had success. The direct correlation between AN and wave parameters (e.g. FM95) is difficult to interpret, because much of that AN-to-wave correlation is via mutual correlation with wind speed, which is already used in the AN predictive model. However, it is possible to anticipate the potential for wave parameters to improve the prediction by performing analysis while controlling for this mutual correlation[6]. We have done this using methods described in the two sections following.

### A. Band-normalized wave parameters

Figure 3 shows examples from the OSP observations at a middle frequency, 841 Hz. The color scale here is the wave parameter $m_3$, normalized by the mean $m_3$ within a wind speed bin of width 1 m/s. Horizontal black lines indicate these bins, and the mean AN in the bin, and thus these black lines can be taken as an example of an AN prediction that is based on wind speed alone. Below we discuss a "wind speed only" model, which refers to a hypothetical step-function empirical model which is composed of these black lines. Fig. 4 show results for the Clayoquot (ONC) model output at two frequencies, 841 Hz and 4700 Hz. Since this case has larger population (10,900 vs. 4400), plotting individual points would result in many concealed points, so instead, the normalized $m_3$ values are plotted as the averages on a 1 m/s × 2 dB grid.

The $m_3$ parameter was selected for this example based on analysis that what we will subsequently present in Section IV.B and Fig. 6. It is not the only wave parameter which shows the trend shown in Fig. 3. The parameters $m_2$, $m_4$, $m_5$, and $D_{tot}$—and to a lesser extent, $m_0$ and $m_1$—also show this. Similar example plots, for $m_2$, $m_4$, $m_5$, and $D_{tot}$ are provided in the Supplementary Material. Examples for a higher frequency, 7.2 kHz, are also provided in the Supplementary Material.

Figure 3 shows that at higher wind speeds, there is little variation of AN from the black horizontal lines. This indicates that a "wind speed only" prediction of AN is already quite accurate, and little can be done to improve upon it. At lower wind speeds, there is much more scatter above and below the black lines. The scatter is presumably due to sources other than wind. The colors indicate that where observed AN is higher than the "wind speed only" model, normalized $m_3$ tends to be higher (red and orange), and below the line, the opposite is true (blue colors). This suggests that some improvement can be made on the "wind speed only" model by incorporating $m_3$ into the prediction model.

Figures 3 and 4 indicate a dependence of AN on wind/wave parameter even at low wind speeds. This may seem surprising given that whitecaps typically start to form only at wind speeds of 4 to 6 m/s (e.g., Monahan 1971): 5 m/s is sometimes used as a rule of thumb. A study by Updegraff and Anderson (1991) sheds light on this, and here we rely on a summary given in Callaghan (2018), in context of so-called microscale breaking at low wind speeds (emphasis added): "There is *acoustic* evidence that small oceanic breaking waves occurring at wind speeds as low as 1.5 m/s may entrain small amounts of air but do not form distinctive whitecaps (Updegraff and Anderson 1991)."

---

[6] Hwang and Sletten (2008) made a similar observation when investigating the dependence of whitecap coverage on wave state: "The weak signal of explicit wave dependence can be detected when the whitecap coverage is divided by the wind speed cubed." In other words, dependence on wind speed dominates, and the problem can be clarified by extracting this dependence and studying the residual.

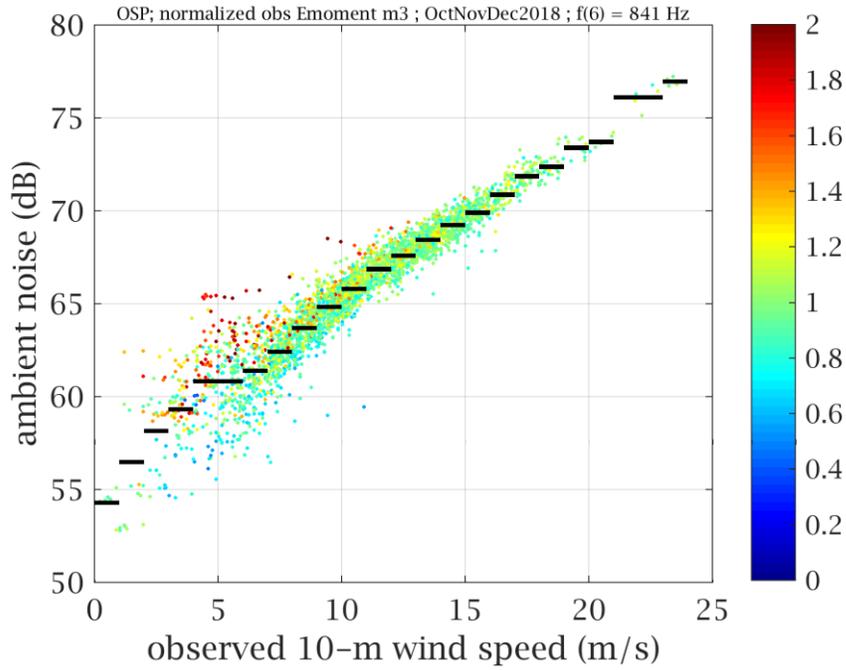

FIG. 3. AN, in dB re: µPa²/Hz, at 841 Hz as a function of wind speed $U_{10}$; colors indicate the value of normalized $m_3$ as described in the text. Results for the OSP case, with $m_3$ computed from observations.

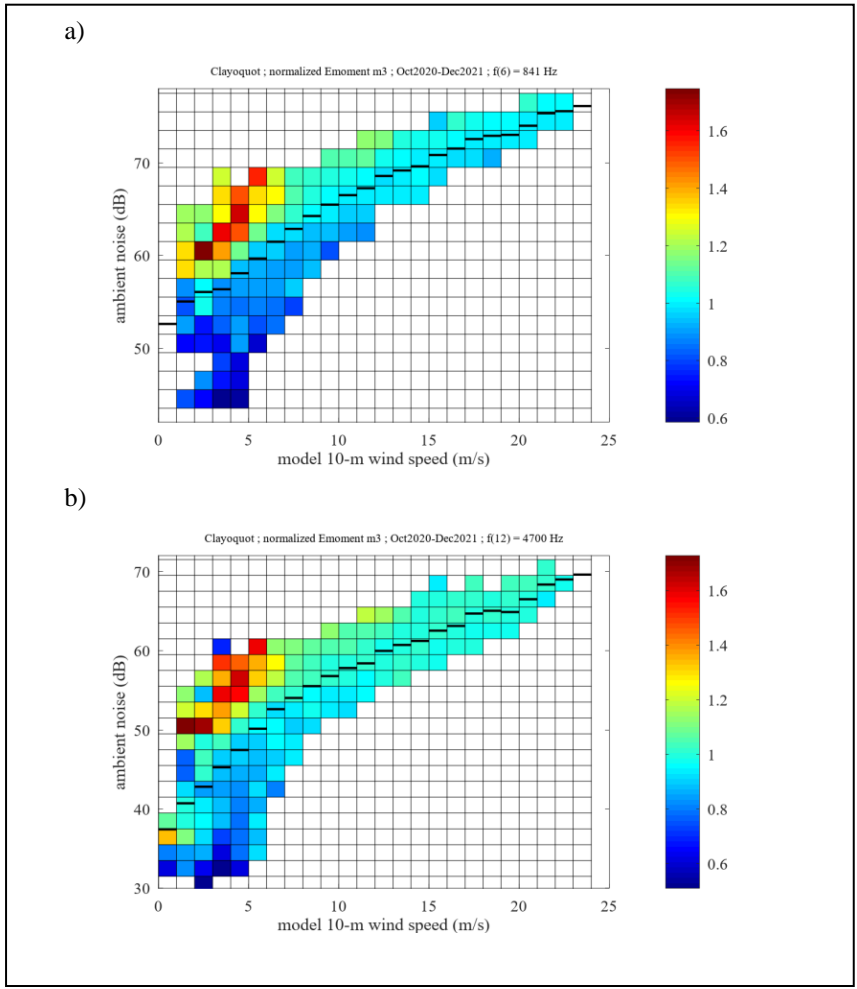

FIG. 4. Like FIG. 3, but showing results for the Clayoquot case, with $m_3$ computed from the ocean wave model, SWAN. Panel (a) shows results at 841 Hz and panel (b) shows results at 4700 Hz. Values of normalized $m_3$ are indicated with the color scale, plotted on a 1 m/s $\times$ 2 dB grid. Where the population of points within a grid cell is less than five, the grid cell is shown as white.

## B. Correlation of residuals

A limitation of the diagrams of Fig. 3 and 4 is that it only presents results for a single wave parameter and single acoustic frequency at a time. Here, we introduce a method which permits averaging over frequencies. First, we group our dataset into 1 m/s bins, essentially looking at vertical slices of plots such as Fig. 4, and then if the population of a bin is not small—we use a threshold of 200 here—we compute the Pearson correlation between AN and the wind/wave parameter. Since correlation with $U_{10}$ is constrained via the binning, we refer to this computed correlation as "residual correlation". Examples for $U_{10}$ and $m_4$ are shown in Fig. 5a, for the wind speed bin of 4 to 5 m/s. Importantly, there is still some correlation between AN and wind speed within this 1 m/s bin: $R_P$=0.10. This establishes a baseline against which to judge the correlation of AN with the wave parameter. We find a correlation for normalized $m_4$ of $R_P$=0.46. The substantial exceedence of the wave parameter $R_P$ over the wind $R_P$ can be taken as an indication that there is correlation of AN with the wave parameter *independent of their mutual correlation with wind speed*. However, this result is not seen across all wind speeds. In Fig. 5b, we plot the correlation values as a function of wind speed and, consistent with what we saw in Fig. 3 and 4, the $R_P$ of the wave parameter exceeds that of the wind speed only at lower wind speeds ($U_{10}$ from 2 to 7 m/s).

The correlation functions $R_P(U_{10})$ at individual frequencies such as Fig. 5b can then be averaged to represent all frequencies in a single plot: Fig. 6 shows the mean magnitude of $R_P(U_{10})$ for several wind/wave parameters. The result for wind speed is included as a thick blue line, to indicate the range in which wave parameter correlation can be considered insignificant (equal to or smaller than thick blue line). This comparison includes two parameters which have not been previously introduced, summarized here: "STEEP" is the mean steepness $H_{sig}/L$, where $L$ is mean wavelength, and "precip" is mean precipitation rate. The parameters are described in detail in Appendix A.

This figure indicates that parameters such as $m_2$, $m_3$, $m_4$, mean steepness, and integrated dissipation ($D_{tot}$) can help to improve the AN prediction, by a modest amount (correlation from 0.25 to 0.42), at lower wind speeds (2 to 7 m/s). We find that despite the effort to remove precipitation during the acoustic data processing stage (see Appendix B), there is significant correlation with precipitation, even at higher wind speeds.

In the moderate to high wind speeds, 10 to 15 m/s, $D_{tot}$ is the only wave parameter in Fig. 6 which has correlation $R_P$ that is elevated relative to that of wind speed, and the difference is slight. Taking example values that are used in the calculations of the average of Fig. 6: for the 10-11 m/s bin, at 841 Hz, the $R_P$ for $D_{tot}$ and $U_{10}$ are 0.23 and 0.14 respectively. At 1.7 kHz, it is 0.22 and 0.14 respectively. This suggests that $D_{tot}$ has only slight potential for improving AN prediction in this wind speed range.

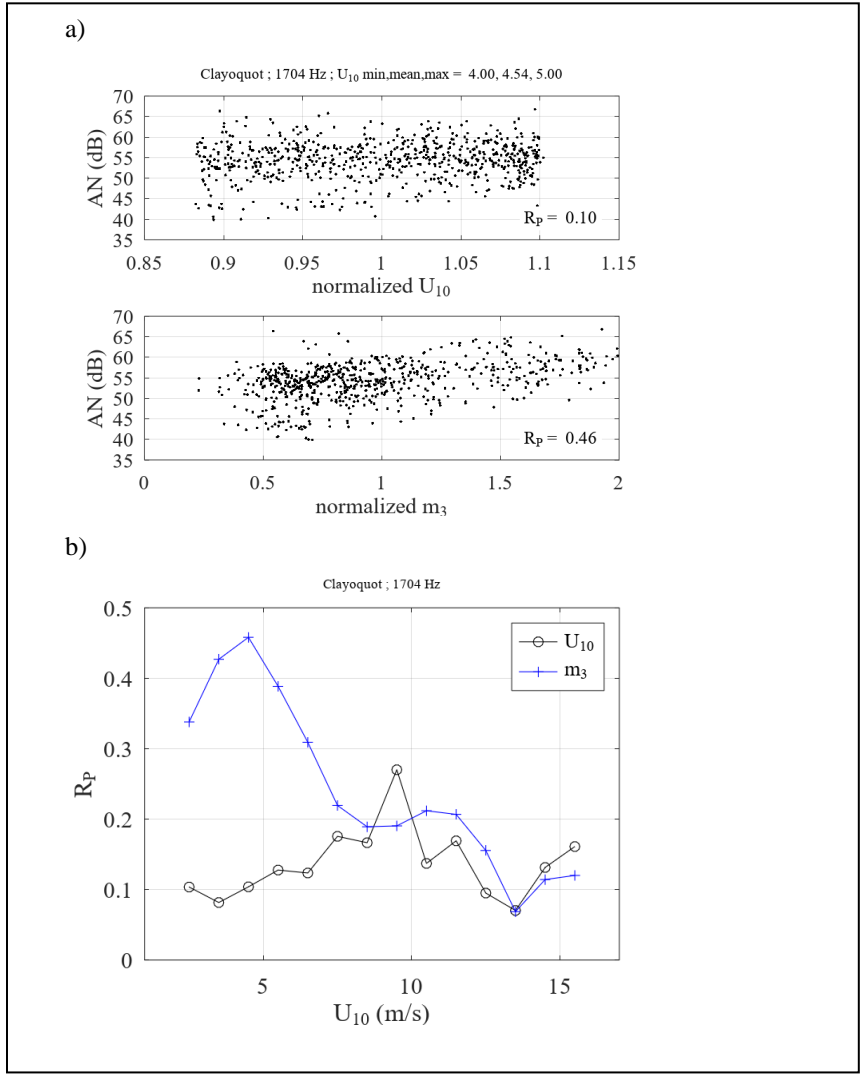

FIG. 5. Correlation calculations for the Clayoquot case at 1.7 kHz. In (a), Results at a single wind speed bin ($U_{10}$=4 to 5 m/s), top panel: AN vs. normalized wind speed $U_{10}$ ($R_P$=0.10), lower panel: AN vs. normalized wave moment $m_3$ ($R_P$=0.46). In (b), dependency of correlation $R_P$ on $U_{10}$ and $m_4$ is shown for all wind speed bins. The plot terminates at $U_{10}$=15-16 m/s, because correlation is not calculated for wind speed bins with population less than 200.

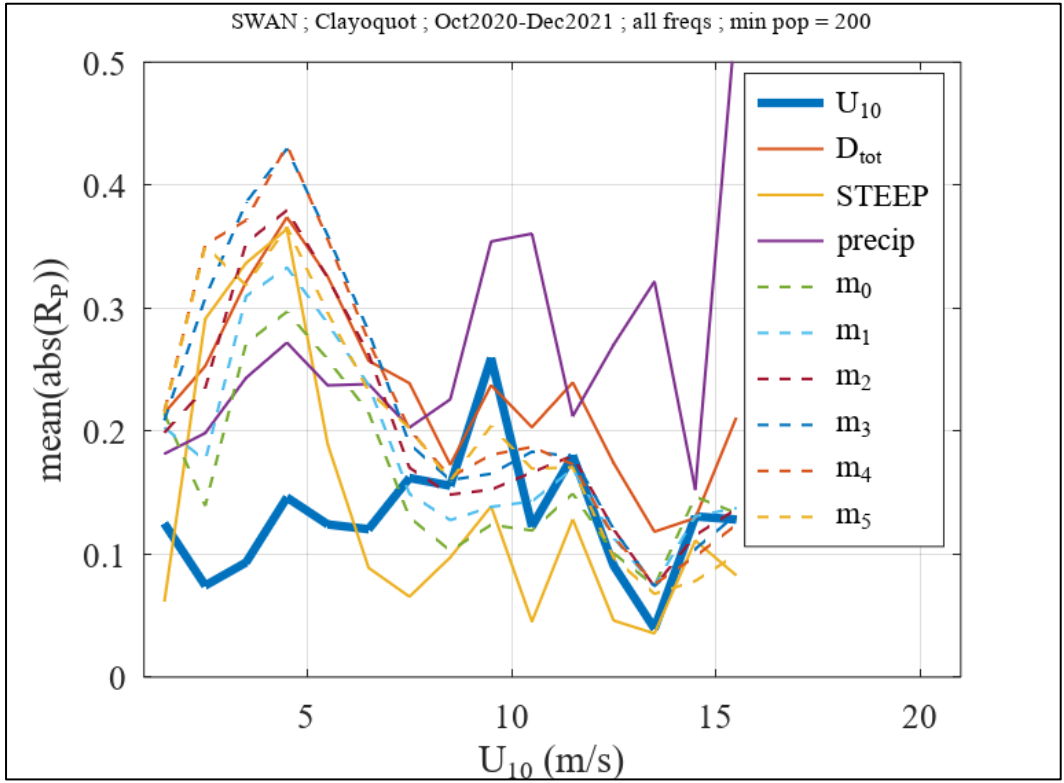

FIG. 6. Clayoquot case: Mean magnitude of the Pearson correlation coefficient $R_P$ as a function of wind speed, for different wind/wave parameters. This is the "residual correlation" remaining after the correlation with wind speed has been constrained (see text).

## V. MULTI-LINEAR REGRESSION

For each location, we perform linear regression to fit individual wind/wave parameters to AN and multi-linear regression to fit multiple parameters (e.g. wind speed plus one to five wave parameters). For each regression, using the notation that $C_1$ is the first in a series of fitting coefficients, and $p_1$ is the first in a series of $n$ input parameters (e.g., wind speed), we create both a linear fit, $AN_{lin} = \sum_{i=1}^{n} C_n p_n = C_1 p_1 + C_2 p_2 + ...$, and a log fit, $AN_{log} = \sum_{i=1}^{n} C_n \log_{10} p_n = C_1 \log_{10} p_1 + C_2 \log_{10} p_2 + ...$, and select the one with higher skill. We also experimented with power fits, but those results are not presented here. The parameter sets included here have $n$ as low as $n = 1$ (the case with $p_1 = U_{10}$), and as high as $n = 8$ (the case with $p = \langle U_{10}, m_{-1}, m_0, \dots, m_5 \rangle$). For example, in Section V.A, we have seven parameters sets, and with 18 frequency bands, this means that for every hydrophone location, 126 fits are recorded, each being of type $AN_{lin}$ or $AN_{log}$.

Multi-linear regression was performed for all seven locations, but the OSP results are excluded here. This is for two reasons. Firstly, the frequency range (127 Hz to 1.7 kHz) is smaller than that from the other cases (127 Hz to 18 kHz), and secondly, AN levels at the lowest frequencies at the OSP location are elevated relative to levels anticipated by models such as Wenz (1962) and Pflug (2021), perhaps from ubiquitous, distant shipping. Both factors complicate the presentation of MLR results from OSP alongside the other cases, in particular the "cross-application" section below.

Because OSP is excluded, all MLR analysis herein is performed using modeled wind and wave parameters.

Below, we present results of multilinear regression (MLR) using two broad approaches. In the first approach, that of "<u>local application</u>", we calibrate using a data set from a location and then apply the MLR model to independent data from the same location. Each AN time series is segmented by "day of month", with a separator at "day of month" equal to 21.0, so that roughly 2/3 of data are used for creating the regression, with 1/3 withheld and subsequently used for prediction and evaluation.

In the second approach, that of "<u>cross-application</u>", we calibrate MLR models for each location individually and apply the models at all six locations, including the location used for the calibration, resulting in 36 calibration/application pairs for each model type. No data are withheld.

Hildebrand et al. (2021) use an explicit formula for sound absorption, and Wenz (1962) divides data into two water depth categories, deep and shallow. Since our prediction models do not control for sensor depth, we should anticipate that they suffer a penalty to overall accuracy. However, since the penalty applies to both the baseline model (using $U_{10}$ alone) and the experimental models (using $U_{10}$ plus wave—and in some cases precipitation—parameters), general conclusions about relative skill are still valid. Unlike those earlier works, our goal here is *not* to create a predictive model that can be applied to other locations. In both cases (local application and cross-application) the goal is the same: *to quantify the impact of inclusion of wave parameters on model predictive skill.*

## A. Local application

Figure 7 shows evaluations of the multi-linear predictions, comparing the "wind speed only" model against six examples of experimental models. The vertical axis is the RMS error of the AN prediction, in dB re: μPa²/Hz. The plot includes two wave parameters that have not yet been introduced: "FSPR" is a metric for frequency narrowness of the wave spectrum and "DSPR" is a metric for directional spread of the wave spectrum. The parameters are described in detail in Appendix A.

The wave parameter set of spectral moments $\langle m_{-1}, m_0, \ldots, m_5 \rangle$ was chosen as a list that we expected to adequately describe spectral shape. The parameters sets with $U_{10}$ plus three other parameters were selected based on early experiments with the MLR, in which the best performing first parameter (which was $m_5$) provided a starting point, and then the most complementary second parameter was chosen ($m_{-2}$), and then several of the most complementary third parameters (FSPR, DSPR, STEEP, $D_{tot}$, precip) were added. Here, "most complementary" refers to RMSE reduction for that preliminary dataset, and we expect that the outcome was determined, at least in part, by the orthogonality of the parameters.

We find that there is modest but consistent improvement in the predictive skill, relative to the baseline model. The degree of improvement is mostly insensitive to the selection of wave parameters.

Some conclusions and conjecture can be made from the results shown in Fig. 7:

- The improvements vs. the baseline model are typically less than 0.3 dB RMSE.
- Improvements are typically small at the lowest and highest frequencies, and largest at the middle frequencies. This implies that our dataset is well-bounded for our problem, i.e., extending the experiment to lower or higher frequencies is not justified.
- The start and end of this mid-frequency range varies from one location to another.
- The selection of additional parameters has only minor impact, i.e., it appears that one set of wave parameters is as good as another.

- The 'Antarc' case is somewhat of an outlier: the baseline error is relatively high (over 4 dB at 2.5 to 6 kHz), as is the degree of improvement (more than 0.7 dB at 2 - 4 kHz). This result is not surprising given that the region of the hydrophone has sea ice present during approximately 50% of the time period in question. While the baseline model does not account for the impact of sea ice on "wind noise", the wave model does.

Though not shown here, we also compared the accuracy of models with just two input parameters, e.g. $U_{10}$ and $D_{tot}$. As a general rule, these were slightly less accurate than the models shown in Fig. 7, which use four or more input parameters.

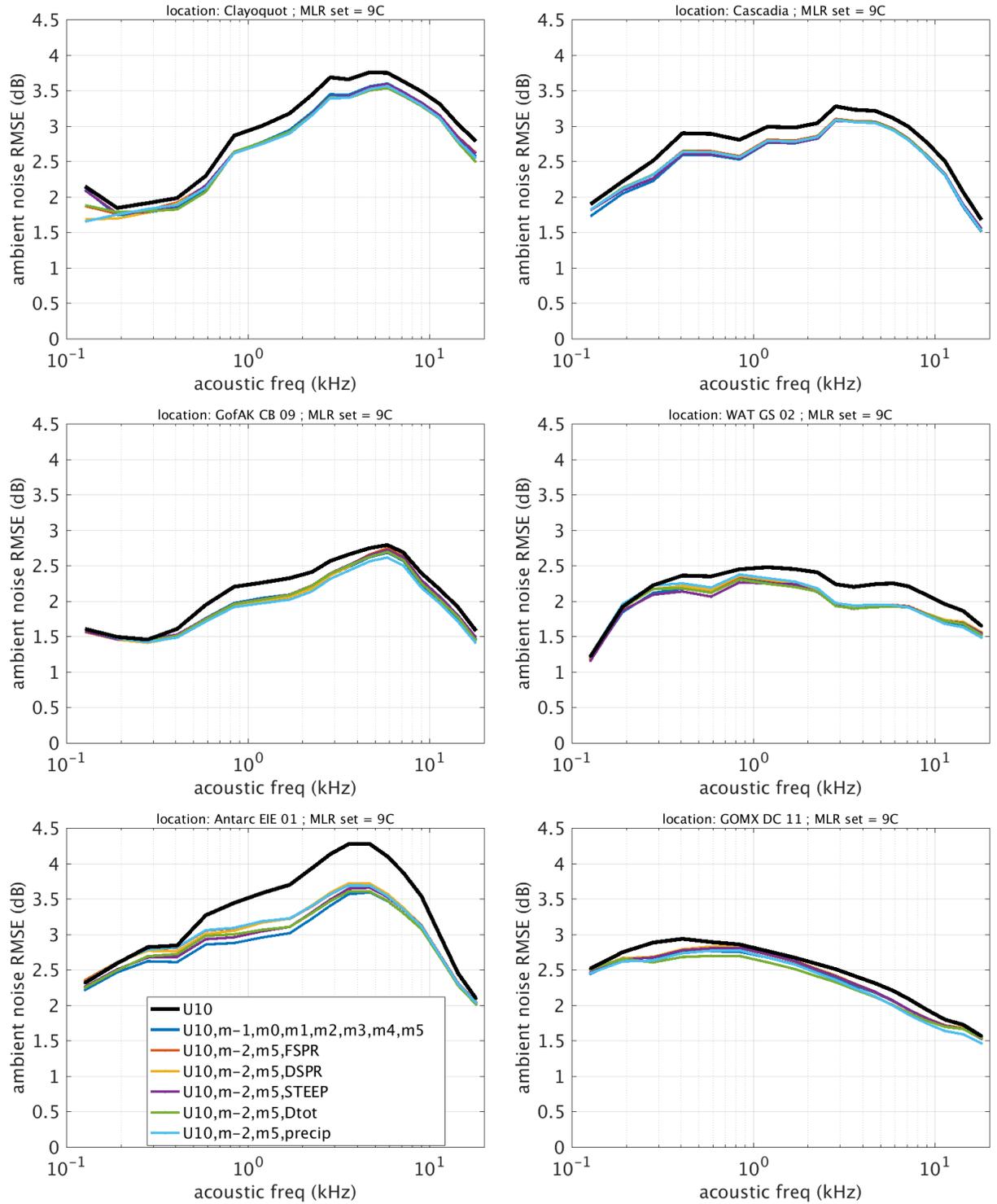

FIG. 7. RMS error in dB re: μPa²/Hz for prediction of AN for the six ONC and HARP locations, using different combinations of wind/wave parameters as inputs to the AN prediction, which is based on multi-linear regression.

## B. Cross-application

With the cross-application experiments, there are 36 combinations per model, making it impractical to present in the format of Fig. 7. Instead, we use tabulations of RMSE and RMSE averages.

Table IV presents the RMSE for each combination for a single frequency bin near the middle of the frequencies used here, 2.85 kHz, and showing a single experimental model, the one using the parameter set $\langle U_{10}, D_{tot} \rangle$.[7] The table indicates that the experimental model is an improvement over the baseline model using parameter set $\langle U_{10} \rangle$ for 29 of the 36 combinations. In the case of the baseline model, the least successful calibration is that which was made using the 'Antarc' case, with a mean RMSE of 3.71 dB. This is unsurprising, since, as mentioned already, the baseline model does not take into account the effect of sea ice on 'wind noise'. The experimental model $\langle U_{10}, D_{tot} \rangle$ calibrated to 'Antarc' outperforms the baseline model $\langle U_{10} \rangle$ calibrated to 'Antarc' for all application cases (mean RMSE of 3.29 dB, reduced by 11.3% from the RMSE of the baseline). Of the experimental model calibrations, the calibration to the 'WAT' case is most successful (mean RMSE of 3.14 dB). For cases other than 'Antarc', improvement is marginal: 0.15 to 0.19 dB from a baseline of 3.32-3.71, or a 4% to 5% improvement from the baseline.

Table V shows results for all frequencies, for the same two models ($\langle U_{10} \rangle$ baseline and $\langle U_{10}, D_{tot} \rangle$ experimental). We see that the experimental model is more accurate than the baseline for 16 out of 18 frequencies, and for all frequencies $\geq 408$ Hz. To show results from several experimental models, we must further aggregate the outcomes. This is done in Table VI, by indicating the number of frequency bins for which the experimental model outperforms the baseline model, using the mean across all 36 calibration-application combinations. Here we find that the model with only two inputs, $\langle U_{10}, D_{tot} \rangle$, out-performs the same model with a third input parameter ($m_3$, $m_4$, $m_5$, or DSPR), which differs from our conclusions from the "local application" study. This may imply that the models with several input parameters are "over-tuned" to their specific locations.

---

[7] The frequency 2.85 kHz is selected based on differences listed in Table V, as having a difference slightly above the median of those listed. The experimental model which uses $U_{10}$ and $D_{tot}$ is selected based on the relatively good results that we will show in Table VI using this parameter set.

Table IV. RMSE (in dB) when applying MLR calibrations to locations which include those different from the location of calibration (see "Formula from" and "Applied to"). Results for the 2.848 kHz band are shown. For each location, there are two rows: the first (lighter shading) shows the RMSE using the regression model with $\langle U_{10} \rangle$ as the only input (baseline model), and the second (darker shading) shows the RMSE using MLR with $\langle U_{10}, D_{tot} \rangle$ as input. The final column indicates the average across the row. The "grand average" (mean of mean) for the baseline model (MLR using only $\langle U_{10} \rangle$) is 3.49 dB and that of the $\langle U_{10}, D_{tot} \rangle$ model is 3.27 dB.

| | Applied to: | | | | | | |
|---|---|---|---|---|---|---|---|
| **Formula from:** | Clayo-quot | Cas-cadia | GofAK | WAT | Antarc | GofMX | mean |
| **Clayoquot** | 3.71 | 3.36 | 2.98 | 2.47 | 4.78 | 3.02 | 3.39 |
| | 3.50 | 3.29 | 2.96 | 2.59 | 4.11 | 2.81 | 3.21 |
| **Cascadia** | 3.84 | 3.21 | 3.47 | 3.01 | 5.39 | 3.35 | 3.71 |
| | 3.67 | 3.11 | 3.49 | 3.13 | 4.72 | 3.22 | 3.56 |
| **GofAK** | 3.90 | 3.86 | 2.70 | 2.26 | 4.19 | 3.15 | 3.34 |
| | 3.83 | 4.03 | 2.47 | 2.13 | 3.56 | 2.94 | 3.16 |
| **WAT** | 3.83 | 3.70 | 2.72 | 2.24 | 4.30 | 3.13 | 3.32 |
| | 3.76 | 3.86 | 2.52 | 2.10 | 3.64 | 3.00 | 3.14 |
| **Antarc** | 4.40 | 4.65 | 2.97 | 2.77 | 3.98 | 3.50 | 3.71 |
| | 4.04 | 4.28 | 2.55 | 2.29 | 3.46 | 3.12 | 3.29 |
| **GofMX** | 3.80 | 3.55 | 3.10 | 2.59 | 4.82 | 2.90 | 3.46 |
| | 3.63 | 3.57 | 2.99 | 2.68 | 4.14 | 2.63 | 3.27 |

Table V. The "grand average" (mean of mean RMSE, in dB re: $\mu Pa^2/Hz$, for all locations), showing results from all 18 frequency bands, for the experiment of applying MLR calibrations to locations which include those different from the location of calibration. The frequency highlighted in gray is the frequency in Table IV (2.848 kHz). As with Table IV, the first model is the regression model which uses $\langle U_{10} \rangle$ as input, and the second is the MLR model which uses $\langle U_{10}, D_{tot} \rangle$ as input. The difference is the baseline model RMSE minus the experimental model RMSE.

| freq. (kHz) | RMSE (dB) model= $U_{10}$ | RMSE (dB) model= $[U_{10}, D_{tot}]$ | difference (dB) |
|---|---|---|---|
| 0.127 | 2.69 | 2.74 | −0.05 |
| 0.191 | 2.76 | 2.75 | 0.00 |
| 0.281 | 2.98 | 2.99 | −0.01 |
| 0.408 | 3.08 | 3.08 | 0.01 |
| 0.587 | 3.08 | 2.99 | 0.10 |
| 0.841 | 3.15 | 3.02 | 0.13 |
| 1.199 | 3.19 | 3.03 | 0.16 |
| 1.704 | 3.21 | 3.02 | 0.20 |
| 2.260 | 3.26 | 3.04 | 0.22 |
| 2.848 | 3.49 | 3.27 | 0.22 |
| 3.588 | 3.35 | 3.11 | 0.24 |
| 4.700 | 3.30 | 3.03 | 0.27 |
| 5.875 | 3.30 | 3.02 | 0.27 |
| 7.175 | 3.35 | 3.08 | 0.27 |
| 9.040 | 3.41 | 3.16 | 0.25 |
| 11.390 | 3.53 | 3.33 | 0.20 |
| 14.350 | 4.00 | 3.86 | 0.14 |
| 18.080 | 4.03 | 3.93 | 0.10 |

Table VI. Summary results of different models, for the experiment of applying MLR calibrations to locations which include those different from the location of calibration. The baseline model is the regression model which use $\langle U_{10} \rangle$ as input. Seven alternative models using MLR are shown, taking inputs as indicated in the second column. A "positive outcome" is a case where the "grand average" (mean of mean) RMSE is lower than that of the baseline model (model #1).

| model # | parameter list (inputs to MLR model for prediction of AN) | # of frequencies with positive outcomes (out of 18) | range of frequencies w/ positive outcome |
|---|---|---|---|
| 2 | $U_{10}, m_3, D_{tot}$ | 7 | 2.26 – 9.04 kHz |
| 3 | $U_{10}, m_4, D_{tot}$ | 6 | 2.85 – 9.04 kHz |
| 4 | $U_{10}, m_5, D_{tot}$ | 13 | 0.84 – 18.08 kHz |
| 5 | $U_{10}, D_{tot}$ | 16 | 0.41 – 18.08 kHz |
| 6 | $U_{10}, D_{tot}$, precip | 16 | 0.41 – 18.08 kHz |
| 7 | $U_{10}$, DSPR, $D_{tot}$ | 14 | 0.59 – 18.08 kHz |
| 8 | $U_{10}, m_3, m_5$, DSPR, $D_{tot}$, precip | 9 | 1.20 – 7.18 kHz |

## VI. DISCUSSION

### A. Limitations and future work

Non-local AN generation is not addressed here. All past studies have, to our knowledge, also omitted this. In principle, this can be addressed with an acoustic propagation model, but the calibration process would become prohibitively complex.

The degree of reduction of RMSE in AN predictions by including wave parameters is found to be marginal herein, but may be larger for fetch-constrained cases. With the possible exception of the 'Antarc' case, datasets that we have used here are not severely fetch-constrained by proximity to land. We recommend that additional experiments be performed for cases where the hydrophone is less than 100 km from shore, and off-shore winds climatologically predominant.

We find here that the degree of improvement to model predictive skill by including wave parameters is increased in specific cases: when predicting AN for frequencies mostly falling in the middle of our frequency range, and for regions that are sometimes ice-covered. Further, our correlation study suggests that more improvement can be anticipated for cases of lower winds. AN prediction tools using wave information which target these specific cases may be most fruitful. To this end, it would be worthwhile to experiment with frequency-space fittings at discrete wind speed bands, in lieu of the MLR fittings at discrete frequency bands used herein.

Our MLR models treat the wave parameters as the independent variables, along with $U_{10}$. It may be useful to experiment with instead using the wind-normalized wave parameters, as was done in our correlation study (e.g., Fig. 3).

### B. Non-monotonic behavior

Where both frequency and wind speed are high, it has been shown that there is a relative reduction in the sound levels. Thus AN is non-monotonic in wind speed dependence at high frequencies. The earliest reference to this phenomenon was, to our knowledge, from Farmer and Lemon (1984). They propose that the effect, observed at 14.5 and 25 kHz, is caused by the scattering and absorption of sound by near-surface bubbles from wave breaking, and we find no studies since then which contradict this explanation. In the parametric AN model of Hildebrand et al. (2021), this starts to appear at frequencies above 10 kHz and $U_{10}$ >15 m/s. Yang et al. (2023) find a "fast drop-off of [ambient noise level] for frequency higher than 4 kHz when wind speed exceeds 15 m/s." Inspecting our ONC Clayoqot and HARP GofMX datasets, we find that AN is unambiguously non-monotonic with wind speed at the frequency bands 14 and 18 kHz, for wind speed greater than 15 m/s; these plots are included in the Supplemental Material.

The non-monotonic behavior is a concern in context of our MLR fitting, which, like Wenz (1962), is monotonic with wind speed. The natural outcome is reduced accuracy of the fit for these data points (high frequency, high wind speed). Hildebrand et al. (2021) use more advanced fitting methods, with fitting parameters that are a function of wind speed and frequency, but this is not practical for a regression to multiple wave parameters, so it was never considered for this study. Machine Learning methods may be able to address this problem.

### C. Omitted content

We make the following conclusions which are based on comparisons which are omitted from this manuscript for sake of brevity:

- In our evaluations of RMSE for the OSP case (analogous to Fig. 7), we find significantly higher skill (up to around 0.5 dB) using input from observations (wind only or wind and waves) relative to using model values of the same. This indicates that accuracy (or lack thereof) of the winds is a primary determiner of AN predictive skill for all cases. This comparison is included in the Supplementary Material.
- We found that prediction of AN from wind/wave parameters using Machine Learning (ML) exhibits skill that is not significantly better than that using the simpler method of multilinear regression. However, since there is great variety in ML methods, this is an approach that should be explored further using our dataset.

### D. Wave age

Herein, wave age is used for general description of climatology (Table II), and our thought experiment (Section II) centers on the issue of wave parameter development relative to a fully-developed condition. However, wave age is not included as an input parameter to our MLR models, which may seem counter-intuitive. There are two reasons for this. First, the actual wave age depends on both wave frequency and direction, i.e., it is a "wave age spectrum", $C(f)/Ucos(\theta)$, where $C$ is wave phase velocity and $\theta$ is the angle between the wind and the wave component. One can imagine a careful calculation as an energy-weighted integration of this spectrum; we have not done this. Second, the dependence of ambient noise on wave age should be non-monotonic, making any linear fitting problematic. This was first shown by Dragan-Gorska et al. (2024): they look at the trend for two datasets: one shows no trend ($R = -0.04$) and another shows negative trend ($R = -0.65$). Our assertion for non-monotonic dependence is based on the following. At very small fetch (or very small wave age), the ambient noise generation will also be very small, as the waves are mere ripples. On the other hand, when wave age is large, the waves are swell, which typically do not break in the open ocean, implying that they do not generate significant ambient noise. As AN is small at the two extremes, monotonicity is impossible.

### E. Ambient noise and sea ice

The present study is concerned with the prediction of so-called "wind noise", though other sources contribute (Section I). These sources are not of primary interest here, but there are indirect connections. For example, our data processing (Appendix B) included removal of data contaminated by sources such as precipitation and shipping. Shipping noise is also connected to sea state insofar as most vessels try to avoid heavy seas. In the case of sea ice, we have noted how ice cover affects the AN generated by wave breaking. Ice also generates ambient noise: thermal-induced cracking creates AN across a broad range of frequencies (e.g. 40 to 600 Hz or even broader, see Milne (1972)), while cracking, shearing, and other mechanical effects produce AN at lower frequencies (under 200 Hz, Greening and Zakarauskas (1992), Sheng et al. (2023)). The mechanical effects can be produced by pressure ridges (from horizontal pressure in the ice pack) and motions induced by nonstationary atmospheric pressure. Ridging, in turn, is produced by effects such as ocean currents, wind drag, and wave forcing. In the latter case, as waves enter the Marginal Ice Zone, they are attenuated, and the lost wave momentum is transferred either to the ocean circulation or to the ice. This "push" by the waves on the ice is discussed in Stopa et al. (2018). For further information more generally on ambient noise from sea ice, the reader is referred to Roth et al. (2012), Bazile-Kinda et al. (2013), Geyer et al. (2016), Ozanich et al. (2017), Tian et al. (2019), Wen et al. (2020), Pajala et al. (2021), and Tollefsen et al. (2022).

## VII.   CONCLUSIONS

We conclude the following, based on results presented:

- We used correlation analysis (constraining the correlation of AN with wind speed, to identify correlation with wave parameters that is independent of mutual correlation with wind speed) and one method of AN prediction (multilinear regression (MLR) models). Both methods point to the same conclusion: that prediction of AN can be improved by including wave parameters, but that the degree of improvement is marginal.
- A corollary to above: our experiments robustly indicate that information about AN exists in the ocean wave state that is not contained in the wind speed alone.

- Likely, the process which leads to ambient noise (wave breaking and bubbles) reaches a mature state more quickly than, for example, wave height. This implies that there is less penalty for using "wind only" prediction of AN than for using "wind only" prediction of wave height.
- The correlation analysis indicates that the potential for improving AN prediction using wave parameters is greatest for lower winds (e.g. 2.5-7.0 m/s).
- MLR models are calibrated to specific locations. A consistent improvement via introduction of wave parameters is seen both when applying to the same location (local application) and applying to all locations (cross-application). However, the improvement is marginal. For the local application, typically RMSE is reduced by less than 0.3 dB and/or less than 12% of the original RMSE.
- In the local application of MLR, improvements via addition of wave parameters are typically small at the lowest and highest frequencies, and largest at the middle frequencies (e.g., 0.6 to 6 kHz for the Antarctic location).
- In the local application of MLR, RMSE of the baseline model (using $U_{10}$ only) is especially high for the Antarctic case. We believe that this is because the baseline model does not account for the impact of sea ice on 'wind noise'. Much like fetch, sea ice is a feature of the ocean which can potentially have strong impact on waves and wave breaking, and the improvement of AN prediction via introduction of wave information is therefore relatively high at this location.
- In the cross-application of MLR models, the models using only two or three input parameters are beneficial over a larger frequency interval than one which relies on six input parameters, which suggests a risk of "over-tuning".

## ACKNOWLEDGMENTS

We are grateful that all observational data used in this study is provided freely to the research community, and acknowledge the significant effort and resources required to make this available. We thank NOAA PMEL (R. Dziak and others), NOAA OST (J. Gedamke) and Ocean Networks Canada for their hydrophone datasets; APL/UW (Jim Thomson and others) for the OSP wave buoy dataset; and the Ocean Climate Stations (OCS) Project Office of NOAA/PMEL (Meghan Cronin and others) for the OSP meteorological dataset. The authors gratefully acknowledge Dr. John A. Hildebrand, UCSD, who was generous with both his data and his time, providing copies of the noise recordings from High-frequency Acoustic Recording Package (HARP) data loggers. The HARP datasets were each supported by a different group: GOMRI, PACFLT, SEFSC, and collaboration with Chile. David Wang (NRL 7330) contributed to prior discussions of energy moments. Aubrey Geary (NRL summer intern in 2023) performed Machine Learning experiments. Gregg Jacobs (NRL 7320) provided useful suggestions for visualization of results. Careful reading by anonymous reviewers resulted in significant improvements to the manuscript. This work was funded by the Office of Naval Research through the NRL Core Program, Program Element 0602435N, Work Unit 6C32, Task Area UW-435-025. The NRL Core 6.2 project is "Ambient noise prediction using wave models". This paper is contribution NRL/7320/JA--2024/7 and has been approved for public release.

## AUTHOR DECLARATIONS

### Conflict of Interest

The authors have no conflicts to disclose.

## DATA AVAILABILITY

Data used here are available as "Ambient noise data processed by NRL, co-located with ocean wave data", Mendeley Data, V1, doi: 10.17632/4v72pgswc6.1 . In the case of the HARP datasets, we are able to provide NRL-processed AN data if the requestor has explicit permission from Dr. John Hildebrand, UCSD, who was the provider of the original AN dataset.

**APPENDIX A. Wave Parameters**

Here we provide definitions of the wave parameters used in this manuscript, and where practical, also describe their physical significance. In Table VII, the parameters are given which are derived from moments of the non-directional ocean wave spectrum. Following this table, other wave parameters are described. A short description of the precipitation variable is also given.

Table VII. Description of parameters used. $E(f)$ is the spectral density of the sea surface elevation, with units of m²s, and $f$ is frequency with units Hz.

| Parameter | Remarks | Example |
|---|---|---|
| $m_{-2} = \int E(f)f^{-2}\,df$ | Negative moments such as this are sensitive to the spectral energy level $E(f)$ at lower frequencies. | |
| $m_{-1} = \int E(f)f^{-1}\,df$ | Proportional to energy flux if directional spread is disregarded. Used in a definition of mean wave period ($T_{m,-1,0}$) | $T_{m,-1,0} = \dfrac{m_{-1}}{m_0}$ |
| $m_0 = \int E(f)df$ | Total variance of sea surface elevation. It is proportional to the total energy of the waves, and is used to compute zero-moment wave height $H_{m0}$ which is the spectrum-based definition of significant wave height, $H_s$. | $H_{m0} = 4\sqrt{m_0}$ |
| $m_1 = \int E(f)fdf$ | Proportional to depth-integrated Stokes transport if directional spread is disregarded. Used in a definition of mean wave period ($T_{m,0,1}$) | $T_{m,0,1} = \dfrac{m_0}{m_1}$ |
| $m_2 = \int E(f)f^2\,df$ | Associated with surface orbital velocity. Used in a definition of mean wave period ($T_{m,0,2}$) | $T_{m,0,2} = \sqrt{\dfrac{m_0}{m_2}}$ |
| $m_3 = \int E(f)f^3\,df$ | Proportional to surface Stokes drift if directional spread is disregarded. | |
| $m_4 = \int E(f)f^4\,df$ | Associated with surface acceleration. Proportional to mean square slope in deep water. | |
| $m_5 = \int E(f)f^5\,df$ | Proportional to near-surface vertical shear of Stokes drift if directional spread is disregarded. The spectral parameter $E(f)f^5$ is associated with the Phillips saturation level (Phillips 1966), sometimes taken as a basis for a breaking threshold (Babanin and Young 2005). | |

Other parameters are as follows:

Peak period is defined as $T_p = 1/f_p$. The peak frequency $f_p$ is the frequency $f$ at which spectral density $E(f)$ is maximum.

Integrated dissipation is $D_{tot} = \int S_{ds}(f)df$. This is proportional to the dissipation of wave energy by deep-water breaking (whitecapping). Here, $S_{ds}(f)$ is the wave model source term associated with this dissipation. Units of $S_{ds}(f)$ and $D_{tot}$ are m² and m²Hz, respectively.

Directional spread is 'DSPR'= $\{2(1-m)\}^{0.5}$. This is the frequency-integrated directional spread, computed from the first order frequency-integrated Fourier coefficients $a$ and $b$ of the directional distribution $D(f, \theta)$. The intermediate parameter $m$ is a "centered Fourier coefficient" (Kuik et al. 1988):

$$m = \left( \frac{a^2 + b^2}{m_0^2} \right)^{0.5}$$

$$a = \iint \cos(\theta) \, E(f, \theta) \, df \, d\theta$$

$$b = \iint \sin(\theta) \, E(f, \theta) \, df \, d\theta$$

Here, $E(f, \theta)$ is the directional form of this spectral density, given in units of m²s/rad or m²s/degree, $E(f, \theta) = E(f) \, D(f, \theta)$ and $\int D(f, \theta) d\theta \equiv 1$.

Frequency narrowness/spread parameter with shorthand 'FSPR' is FSPR$= \left| \int E(\omega) e^{i\omega\tau} \, d\omega \right| / m_0$, where $\omega = 2\pi/f$ and $\tau = T_{m,0,2}$. This is taken from Battjes and van Vledder (1984). Contrary to the name, a narrower spectrum implies a higher FSPR (e.g., Rogers and van Vledder 2013), but we retain the notation "FSPR" for consistency with notation of SWAN (2019). A narrower spectrum is generally associated with a more amplitude-modulated or "groupy" wave trains, e.g. Gemmrich and Thomson (2017).

Steepness with shorthand 'STEEP' following SWAN (2019), is computed as $H_{m0}/\lambda_m$, where $\lambda_m$ is the mean wavelength $\lambda_m = 2\pi \left[ \frac{m_0}{\int k E(f) df} \right]$, and $k$ is the wavenumber at frequency $f$, computed from the linear dispersion equation.

Precipitation, with shorthand 'precip' here, is taken from ERA5 reanalysis. This is the mean rate of precipitation. Though the units have no bearing on results presented herein, we apply the MLR using 'precip' with units mm/hr.

**APPENDIX B. Acoustic data processing and quality control**

Calibrated spectral estimates were formed from the raw spectra based on Transmission Voltage Response (TVR) curves for the receivers which accompanied the data sets. The calibrated spectral intensities were linearly averaged over 1-min intervals and band levels were derived as the linear mean of the time averaged spectra for the analysis frequency bands (8 for OSP and 18 for ONC and HARP). For the PMEL data this averaging was done using 1s FFTs with 50% overlap (119 snapshots per minute, 1 Hz frequency resolution) while the ONC and HARP data were processed using 0.1s FFTs with no overlap (600 snapshots per minute, 10 Hz frequency resolution). Finally, an algorithm was developed for automated removal of wind noise contaminants based on statistical techniques for outlier removal. Outlier selection is based on signal normalization techniques and median-based statistical measures; the wind speed is not used directly. These measures include a sliding split window normalizer (Summers 2018) combined with the Median Absolute Deviation (MAD) statistic (Rousseeuw and Croux 1993) and Mean-to-Median Ratio (MMR) statistic (Schwock and Abadi 2021a,b), utilized for threshold-based removal of time-frequency localized interference. For all site data, time domain transient detection was applied to each bin of normalized band averaged results. The MAD statistic was independently computed, in dB, across all relevant frequency bins of the averaged spectra, and the MMR statistic was formed as the ratio (dB difference) of the averaged power spectra and correspondingly computed median spectra. Thresholds applied for each detection criteria were empirically chosen to maximize interference removal while minimizing collateral damage due to the erroneous exclusion of valid wind noise data. Figure 8 shows example spectra from the Clayoquot Slope site along with excluded points (dark blue and magenta bins/regions). Indicated regions of precipitation and shipping noise, along with other sources of interference, were largely removed across all data sets by automated pruning. Figure 9 displays corresponding band wind noise estimates and coincident wind speed (magenta curves).

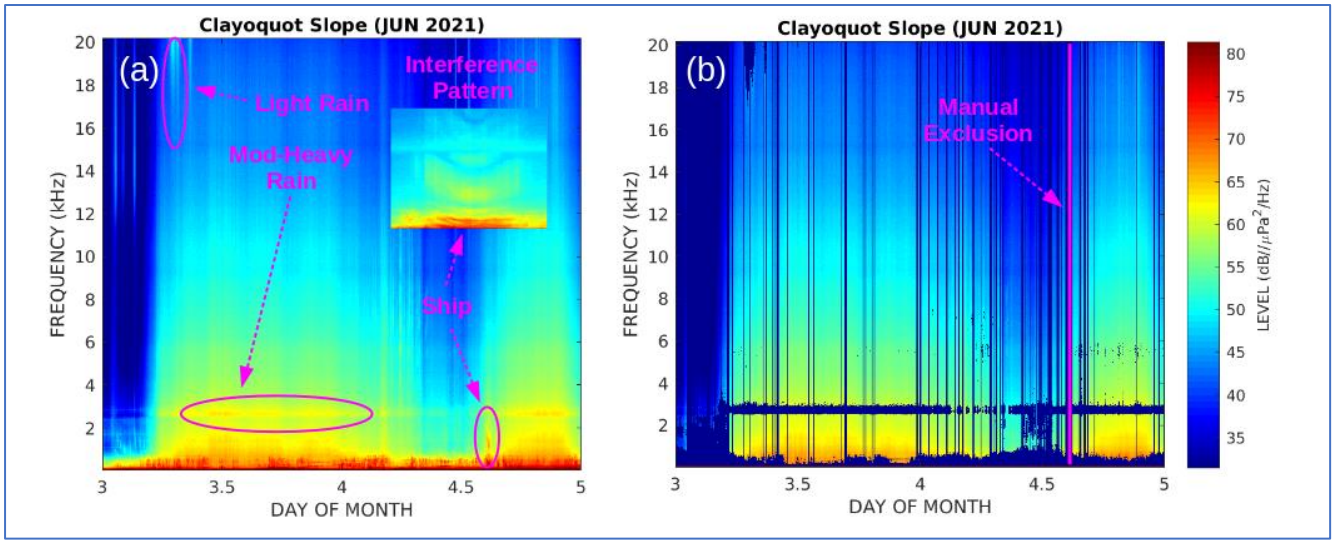

FIG. 8. (a) Clayoquot Slope averaged spectra for days $3 - 4$ of June 2021 with indicated rain and shipping noise, and (b) edited spectra with bins excluded by automated pruning (dark blue) and manual editing (magenta).

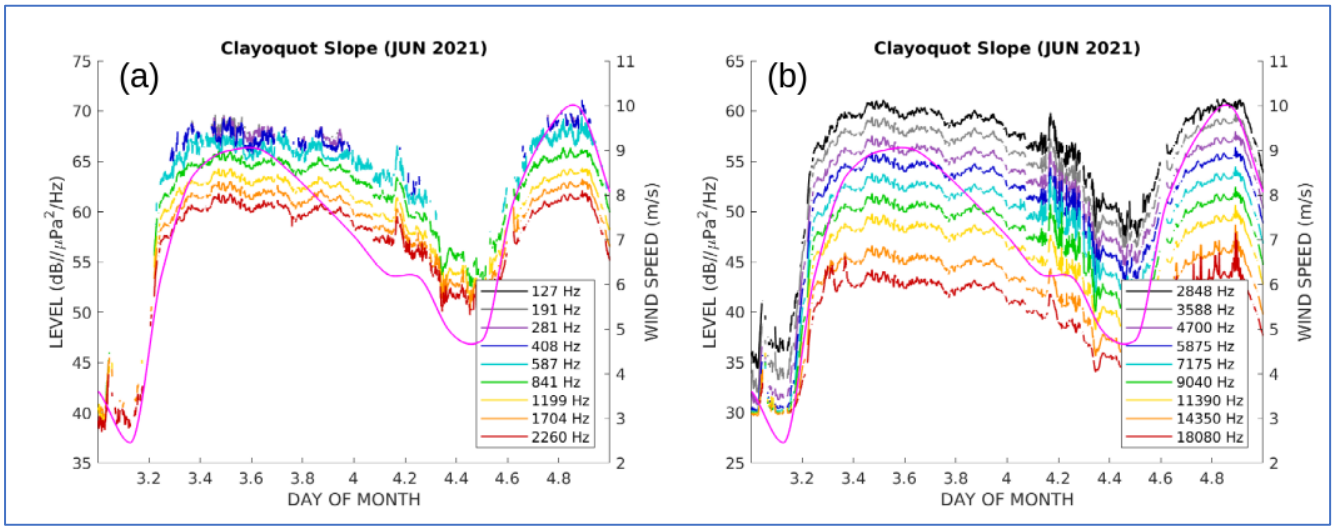

FIG. 9. Band levels derived from the edited spectra in FIG. 8b overlaid with coincident wind speed (magenta curves) for (a) bins $1 - 9$, and (b) bins 10 - 18. Wind speed scale shown on the right of each plot.